\documentclass[preprint,secnumarabic,amssymb, nofootinbib, aps, prd, 11pt]{revtex4-1}
\usepackage{graphicx}				
\usepackage{hyperref}
\usepackage{amssymb}
\usepackage{amstext}
\usepackage{setspace}

\newcommand{\be}{\begin{eqnarray}}
\newcommand{\ee}{\end{eqnarray}}

\newcommand{\qqgn}{g^{n-2k}(\bar{q}q)^k}

\newcommand{\sym}{\mathcal{N}=4 \text{ SYM}}
\newcommand{\ma}{\mathcal{A}}
\newcommand{\mak}{\mathcal{A}^{n_f=k}}
\newcommand{\mao}{\mathcal{A}^{n_f=1}}
\newcommand{\lcl}{(\lambda,\bar{\lambda})}

\begin{document}

\title{Getting more flavour out of one-flavour QCD}%

 \author{Tom Melia}%
\email{thomas.melia@cern.ch}
\affiliation{CERN Theory Division, CH-1211,\\
 Geneva 23, Switzerland.}
\preprint{CERN-PH-TH/2013-290}%

\begin{abstract}{

%
We argue that no notion of flavour is necessary when performing amplitude calculations
in perturbative QCD with massless quarks.
 We show this explicitly at tree-level, using a flavour recursion relation to obtain
  multi-flavoured QCD from one-flavour QCD.
  The method relies on performing a colour decomposition, under which the one-flavour primitive
  amplitudes have a structure which is restricted by planarity and cyclic ordering. 
An understanding of $SU(3)_c$ group theory relations between QCD primitive amplitudes
and their organisation around the concept of a Dyck tree
 is also necessary.
The one-flavour primitive amplitudes
are effectively $\mathcal{N}=1$ supersymmetric, and a simple consequence is
 that all of tree-level massless QCD can be
obtained from Drummond and Henn's closed form solution to tree-level $\mathcal{N}=4$ super Yang-Mills theory. 
}
\end{abstract}

\maketitle
\section{Introduction}
\label{intro}

Fixed-order theoretical predictions for jet cross sections at the Large Hadron Collider (LHC) 
require the calculation of scattering amplitudes in perturbative QCD  involving light QCD partons -- gluons and quarks of
different flavour. For inclusive quantities, all possible partonic contributions must be computed and summed;
for example, for 
6 jet production at tree-level, amplitudes for $gg \to g g g g g g$, $d\bar{d}\to ggggu\bar{u}$, $gg \to u\bar{u}u\bar{u}u\bar{u}$, $gg\to d\bar{d}s\bar{s}c\bar{c}$, {\it etc.}
 are all needed.

What is the difference between the  amplitudes  describing $gg\to u\bar{u}u\bar{u}u\bar{u}$ and 
$gg\to d\bar{d} s\bar{s} c\bar{c}$ in pure QCD, in the limit that all quarks can be treated as massless? 
Since all flavours of quark have the same colour interactions, the only difference comes from the restriction
that flavour conservation imposes on the possible factorisation channels of the amplitudes. 
The amplitudes for the one-flavour case can be obtained from the amplitudes for the distinct-flavour case
 via a permutation sum over quark indices,
\be
\mathcal{M}(gg\to u\bar{u}u\bar{u}u\bar{u})=\sum_{\mathcal{P}(d,s,c)}(-1)^{\text{sgn }\mathcal{P}} \mathcal{M}(gg\to d\bar{d} s\bar{s} c\bar{c})\,,
\label{eq:norm}
\ee
where $\mathcal{P}$ gives permutations of momentum and colour indices of the quarks, 
and the $(-1)^{\text{sgn }\mathcal{P}}$ accounts for Fermi statistics.
For this reason it is taken that the distinct-flavour case is the more general one which should be calculated.

In this paper we show that it  is possible to reverse eq.~\ref{eq:norm} for massless QCD amplitudes at tree-level -- more specifically
for tree-level primitive amplitudes. 
That is, we can recover the distinct-flavour case using primitives involving only one flavour of quark
line. The result follows from an understanding of $SU(3)_c$ group theory relations between the QCD primitives, and a generalisation
of the Dyck basis for quark amplitudes found in Ref.~\cite{Melia:2013bta}. It means that all of tree-level massless
QCD can be rendered effectively $\mathcal{N}=1$ supersymmetric and can as such be obtained from the known solution
to $\sym$ at tree-level.  It is interesting that no notion of flavour is needed at the level of the field theory amplitude calculation (we will
argue this is true at any loop order, using unitarity based methods to obtain loop amplitudes from tree-level amplitudes), 
rather it can be reinstated through combinatorics of the one-flavour tree-level amplitudes alone.

A crucial step is to use a colour decomposition to define primitive amplitudes. 
 By considering a theory where all particles, including
 quarks, are the in adjoint representation, they can be defined at tree-level as \cite{Berends:1987cv,Mangano:1987xk}
 \be
 \mathcal{M}^{\text{tree}}= \sum_{\sigma\in S_{n-1}}\text{tr}(\lambda^1\lambda^{\sigma_1}\ldots \lambda^{\sigma_{n-1}})\,\ma(1\sigma_1\ldots\sigma_{n-1})\,,
 \label{eq:cold}
 \ee 
 where $\lambda^a$ are $SU(3)_c$ fundamental representation matrices, and the $\ma$ are purely kinematic primitive amplitudes.
Since only the colour part of the quark interactions change when putting them into the adjoint representation, the purely kinematic 
primitive amplitudes are the same as in usual QCD, so 
we shall refer to them simply as QCD primitives -- algorithms exist to relate these primitives back to the 
full amplitude with fundamental quarks \cite{Ellis:2011cr, Ita:2011ar,Badger:2012pg, Reuschle:2013qna} 
(at both tree and one-loop level). The primitive amplitudes inherit a number of properties from the colour decomposition
eq.~\ref{eq:cold}.  They are gauge invariant and, with reference to their Feynman diagram representation,
 they only receive contributions from graphs (once drawn in a planar fashion) with
 a cyclic ordering of the external legs that is the same as the labelling of the primitive, giving them a simplified  
 kinematic structure.
The  colour decomposition with all particles in the adjoint representation makes it clear
that general QCD primitive amplitudes satisfy the same group theory relations as do all-gluon amplitudes -- these are known
as Kleiss-Kuijf (KK) relations \cite{Kleiss:1988ne}. However, as was explained in \cite{Melia:2013bta}, the quark lines present in 
the amplitude impart further structure on these relations, modifying them with the effect that fewer than $(n-2)!$ primitives
are independent for $n$ particle scattering. 
In the pure quark and antiquark case considered, with all-distinct flavour quark lines,
the number of independent primitives was shown
to be $(n-2)!/(n/2)!$. In this paper we generalise this statement to QCD 
primitive amplitudes with any number of gluons as well as quarks. Independence for most of this paper will be taken
to mean independent over the field of real numbers (Bern-Carrasco-Johansson (BCJ) relations \cite{Bern:2008qj} 
take into account the further possibility of multiplying primitive amplitudes by kinematic invariants).

In one of the landmarks of the recent progress in the uncovering of the structure of scattering amplitudes, 
 a solution to tree-level $\sym$ was written down by Drummond
and Henn, \cite{Drummond:2008cr}, which provides a closed form for the primitive amplitudes of this theory. 
That such a remarkable formula exists is made possible by the high degree of
symmetry present in the theory. But it has long been understood
that some amplitudes in QCD are effectively supersymmetric at tree-level \cite{Parke:1985pn, Kunszt:1985mg}, and
 in a recent paper, \cite{Dixon:2010ik}, it was shown that all QCD amplitudes with up to four quark lines of distinct flavour
(four also being the number of gluino flavours in $\sym$)
can be obtained from the formula of Drummond and Henn. 
In this paper we show that in fact all of massless QCD can be obtained from $\sym$, even though 
there are only four flavours of gluino in $\sym$ (seemingly less than what is needed to describe QCD) 
as well as possible contributions
from scalar particle exchange (seemingly more than what is needed to describe QCD).

A main motivation for the study of these amplitudes is to be able to make fixed order theoretical 
predictions with which to compare to collider data. The ATLAS and CMS experiments have presented 
data with up to as many as 10 hard QCD jets \cite{Aad:2013wta, CMS:2013gea}, and this presents an enormous challenge for such  computations (at around this number of jets, they are also expected to break down). 
 At leading order (LO) in QCD, calculations with up to 8 jets are available \cite{Gleisberg:2008fv}, and there are
 next-to-leading order (NLO) descriptions of processes involving up to 5 jets \cite{Bern:2013gka,Badger:2013yda}.
QCD tree-level primitive amplitudes
can be considered as fundamental gauge-invariant building blocks of such calculations -- as well
as describing tree-level contributions, they arise naturally in unitarity based 
methods for obtaining the one-loop part of the NLO corrections to  jet cross sections at the LHC \cite{Bern:1994zx,Bern:1997sc,Britto:2004nc,Forde:2007mi, Ossola:2006us,Ellis:2007br,Giele:2008ve,Berger:2008sj} (see e.g. \cite{Ellis:2011cr} for a review),
and also in calculations of QCD amplitudes at higher loop order \cite{Mastrolia:2011pr,Kosower:2011ty,Badger:2012dp,Zhang:2012ce,Badger:2012dv,Johansson:2013sda,Sogaard:2013fpa}. Primitive amplitudes can also be used explicitly 
in subtraction schemes, as formulated within the colourful \cite{Frixione:2011kh}  Frixione-Kunszt-Signer
 framework \cite{Frixione:1995ms}.
  Both a knowledge of a general basis and the flavour recursion described in this paper should be useful for multi-leg
  QCD calculations at leading and next-to-leading order. 
  
  Although for inclusive light jets, multi-quark
 contributions are less important numerically than the more gluonic ones, 
  more exclusive multi-quark final states can be experimentally defined by  tagging the quark-like contributions. 
 Bottom quarks  (and with less efficiency, charm quarks) can be identified by displaced vertices 
   -- 
 as many as 4 $b$-tags are used in ongoing experimental analyses --  and jet substructure 
 techniques \cite{Gallicchio:2011xq, Larkoski:2013eya} can statistically distinguish between jets originating from a light quark and those 
 originating from gluons \cite{ATLAS:2011rga,Chatrchyan:2012sn,CMS:2013kfa,Pandolfi:2013ona}, and are  increasingly 
 being used, for example, in searches for 
 supersymmetric particle hadronic cascade decays. The definition of jet flavour can be made infra-red safe for
 calculations at parton level \cite{Banfi:2006hf}.
    For LHC phenomenology, 
  the addition of electroweak bosons into  amplitudes, as well considering massive top (and possibly bottom) quarks
  is important. 

The outline of this paper is as follows. Sec.~\ref{gen}  describes 
the construction of a general basis of QCD primitive amplitudes for $\qqgn$ scattering
with all-distinct quark flavours, which is of size $(n-2)!/k!$. The notion of a rooted oriented Dyck tree is introduced 
and the connection with quark line structure is made. Sec.~\ref{flav}
presents the flavour recursion, which enables a $k$-flavour primitive to be expressed in terms of one-flavour
primitives, and discusses how $\sym$ amplitudes can be used to completely specify
all of tree-level massless QCD.
We discuss moving away from massless QCD, the role of BCJ relations, and
further directions in Sec.~\ref{disc}, and conclude in Sec.~\ref{conc}.

\section{A general tree-level QCD primitive basis}
\label{gen}

\begin{figure}
\includegraphics[width=17cm]{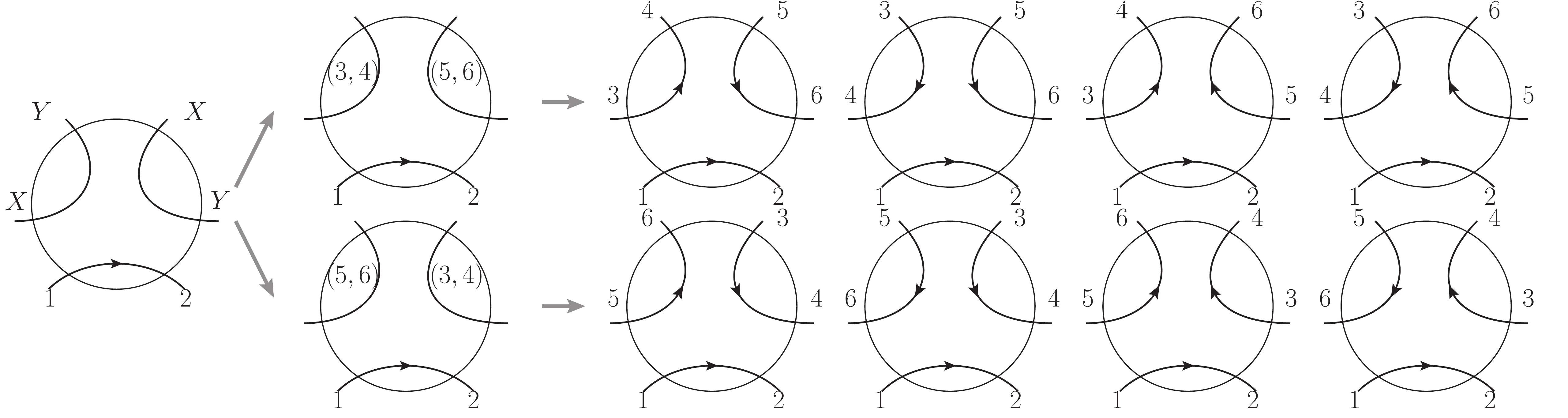}
\includegraphics[width=17cm]{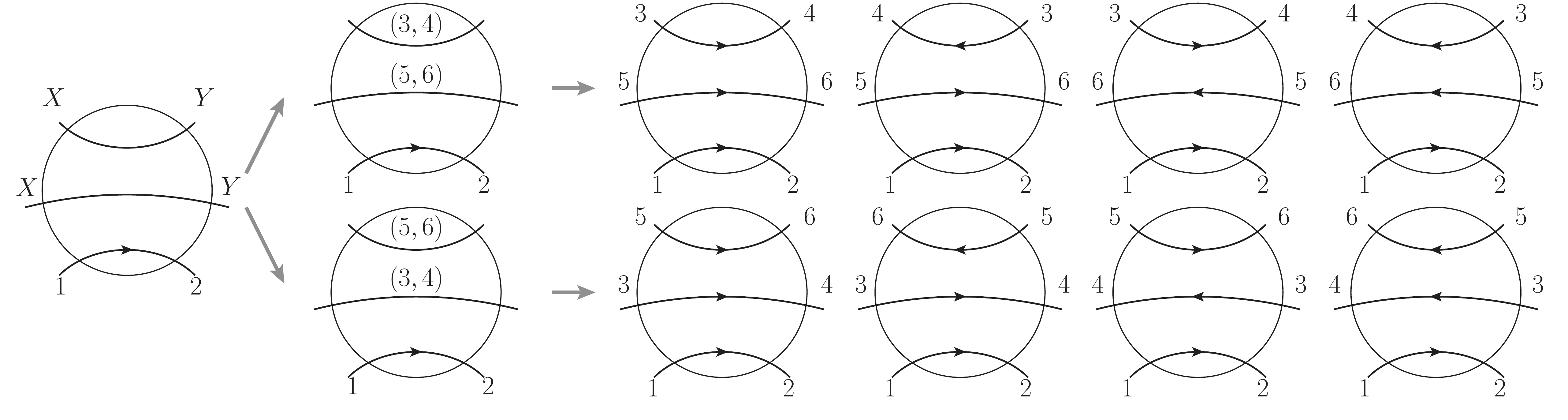}
\caption{Constructing quark line graphs based around the Dyck words $XYXY$ and $XXYY$. The first column shows
the two Dyck topologies, and the second column shows the two possible flavour pair allocations for each topology. Following
the direction of the arrows, each row then depicts the four possible choices of signature for each flavour pair allocation.  
Four permutations corresponding to one graph out of each of these rows constitute the Dyck permutations
 which are then used to construct a basis of QCD primitives.}
\label{fig:6examp}
\end{figure}

The use of Dyck words in understanding the number of independent primitives for purely multi-quark scattering with distinct
flavours of quarks
was presented in \cite{Melia:2013bta}. In this section we will generalise these results to QCD primitives of the form $g^{n-2k}(\bar{q}q)^k$  i.e. with $k$ distinct flavour quark lines and $n-2k$ gluons.
Throughout this section, we will use the convention that the quarks and antiquarks are labelled by the numbers $1\ldots 2k$,
with antiquarks given odd labels, quarks given even labels, and flavour pairs labelled by consecutive numbers: $(1,2)$, $(3,4)$, ..., $(2k-1, 2k)$. The gluons are labelled by the numbers $2k+1\ldots n$.
There is no loss of generality in using this convention for the purposes of this section. 
It will however be necessary to introduce additional notation in the following section to distinguish different flavour pairings. 

We will first state the result: for $k$-flavour quark QCD primitive amplitudes $g^{n-2k}(\bar{q}q)^k$, a basis of primitives is 
the set 
\be
\ma(\,1\, \ldots\, {\sigma_{1}}  \,\ldots\,   {\sigma_{2}}  \,\ldots\,   ~~~  \, \ldots\,  {\sigma_{2k-2}}  \, \ldots  \, 2\,)
\ee
where the $\ldots$ stand for all possible insertions of the gluons $2k+1,\ldots, n$ inbetween the labels 
 $\sigma=\{\sigma_1,\ldots,\sigma_{2k-2}\}$  which are `Dyck permutations' of the set of quark indices $\{ 3,4,5,6,\ldots ,2k-1,2k  \}$. 

The Dyck permutations needed are obtained as  follows (each step is described in detail below): {\it i)} consider all Dyck topologies
arising from Dyck words of length $2k-2$, {\it ii)} for each Dyck topology, consider all possible assignments of flavour pairs, and {\it iii)} for each flavour pair assignment, include only one orientation of each of the flavour pairs.

The Dyck topologies  needed for step {\it i)} are obtained from Dyck words.
Dyck words of length $2k-2$ are strings
of $k-1$ letter $X$s and $k-1$ letter $Y$s with the requirement that the number of $X$s is always greater than
or equal to the number of $Y$s in any initial segment of the string. For each of these words, the Dyck topology
is obtained by identifying pairs of $(X_i,Y_i)$ in the following way: reading the 
word left to right, pair each $Y$ that you come across 
with the closest un-paired $X$ to the left. There are $k-1$ pairs constructed in this way for each Dyck word, labelled with
$i=1,\ldots,k-1$. 

The different flavour pair assignments needed for step {\it ii)} are simply the $(k-1)!$ different possible
ways of assigning the flavour pairs $(3, 4), (5, 6)$ {\it etc.} to the $k-1$ Dyck topology pairs $(X_i,Y_i)$. This is done
for each Dyck topology.

Finally, in step {\it iii)} above, the Dyck permutations are obtained for each of these flavour pair assignments by
 choosing the orientation of the flavour pair.
This means placing the label of each quark and antiquark pair into the Dyck 
word either as $\bar{q}\to X_i$, $q\to Y_i$ or $q\to X_i$, $\bar{q}\to Y_i$, where $(\bar{q},q)$ is whichever flavour pair is assigned to the Dyck topology pair $(X_i,Y_i)$. 
Only one orientation of each quark line is included for each of the flavour assignments arising from the Dyck topologies. We can
introduce the concept of the {\it signature} of the permutation, which is a string of $k-1$ $+/-$ signs depending on whether
$(\bar{q}, q)$ is assigned as $\bar{q}\to X$, $q\to Y$ (a $+$ sign), or $q\to X$, $\bar{q}\to Y$ (a $-$ sign). So we obtain
a vector $(\pm, \pm,\ldots,\pm)$, where the $i$th entry corresponds to the orientation for the pair $(X_i,Y_i)$.
 
 As an example, for $k=3$, there are two Dyck words of length $4\,(=2k-2)$: $XYXY$, $XXYY$. 
  For the first Dyck word the pair assignment is   $X_1 Y_1 X_2 Y_2$.
For the flavour assignments  $(3,4)\to (X_1Y_1),(5,6)\to (X_2Y_2)$,  the permutations
  \be
  (3,4,5,6), (4,3,5,6), (3,4,6,5), (4,3,6,5)\label{eq1} \ee are obtained with signatures $(+,+), (-,+), (+,-), (-,-)$ respectively; from the second possible
  flavour assignment $(5,6)\to (X_1Y_1),(3,4)\to (X_2Y_2)$,  the permutations
  \be
  (5,6,3,4), (6,5,3,4), (5,6,4,3), (6,5,4,3) \label{eq2} \ee are obtained with signatures $(+,+), (-,+), (+,-), (-,-)$ respectively.
  For the second Dyck word the pair assignment is   $X_2 X_1 Y_1 Y_2$.
For the flavour assignments  $(3,4)\to (X_1Y_1),(5,6)\to (X_2Y_2)$,  the permutations
  \be
  (5,3,4,6), (5,4,3,6), (6,3,4,5), (6,4,3,5) \label{eq3} \ee are obtained with signatures $(+,+), (-,+), (+,-), (-,-)$ respectively; from the second possible
  flavour assignment $(5,6)\to (X_1Y_1),(3,4)\to (X_2Y_2)$,  the permutations
  \be
  (3,5,6,4), (3,6,5,4), (4,5,6,3), (4,6,5,3)\label{eq4}  \ee are obtained with signatures $(+,+), (-,+), (+,-), (-,-)$ respectively. A basis
  then consists of four permutations, one chosen 
  from each of eqs.~\ref{eq1}-\ref{eq4}. The four permutations do not have to have the same signature as each other.

A diagram showing the quark line structure of the primitive can be drawn for each of these permutations, see Fig.~\ref{fig:6examp}. 
A quark line is drawn
between each of the identified pairs $(X_i,Y_i)$. The $(k-1)!$ flavour pair allocations change the flavour of these quark lines, and the 
$2^{k-1}$ different signatures for each of these allocations are all possible ways of drawing the direction of the arrow
 on each of the quark lines. Gluons can be inserted in any positions between the quark lines, so at all points
 around the circle, except between 1 and 2 which have been fixed using the KK relations to be always consecutive, as described below. 
 These gluons are not drawn on the quark line graphs. A particularly 
 useful feature of these graphs is that if the quark lines cross then the primitive has to vanish, since with only
 planar contributions, the quark lines are forced to intersect, which results in a flavour-violating vertex.
 
 This is a generalisation of what was found in \cite{Melia:2013bta} in two directions. There is the addition of gluons, and also the
 freedom in choosing the signature of the Dyck permutations describing the quark line structure. 
 We now discuss
  each of these generalisations in turn.
 
 \subsection{Addition of gluons}
 Considering first just the quark line structure, the number of permutations $\sigma$ is given by the number of Dyck words of 
 length $2(k-1)$, which is given by the Catalan number $C_{r}=(2 r)!/r!(r+1)!$ with $r=k-1$, multiplied by the number of
 flavour allocations, $(k-1)!$. 
 Next, the counting of all possible insertions of the gluons can be done by considering all possible distributions of
 an ordered set of the $p=n-2k$ gluons into $2k-1$ slots between the quark lines (excluding between $(1\to2)$), which is given by
a binomial coefficient $(2k-2 +p)!/(2k-2)!/p!$, and then considering all ordered sets, given by the $p!$ possible gluon permutations.
That is, the number of independent $k$-flavour $g^{n-2k}(\bar{q}q)^k$ QCD primitive amplitudes is
 \be
 (\#\text{Dyck words})\cdot (\#\text{Flavour pair allocations})\cdot (\#\text{Gluon distributions})\cdot (\#\text{Gluon permutations}) \nonumber \\
=\frac{(2k-2)!}{(k-1)!k!}\cdot (k-1)!\cdot \frac{(p+2k-2)!}{(2k-2)!p!}\cdot p!~~~~~~~~~~~~~~~~~~~~~~~~~~~~~\, \,\nonumber \\
= \frac{(p+2k-2)!}{k!}=\frac{(n-2)!}{k!}\,.\,~~~~~~~~~~~~~~~~~~~~~~~~~~~~~~~~~~~~~~~~~~~~~  
 \ee
This generalisation coming from the addition of gluons to amplitude is straightforward, since they impart no further structure
which can affect KK relations further than beyond the impact of the quark line structure. The KK relations allow
for two labels to be fixed in consecutive order -- as in Ref.~\cite{Melia:2013bta} we chose here to fix 1 and 2 to be cyclically next to
each other (i.e. we always consider primitive amplitudes of the form $\ma(1\ldots2)$). 
All of the further information concerning relations between the remaining $(n-2)!$ primitives (after fixing 1 and 2)
is contained in the quark line structure, to which we now turn. 

\subsection{A rooted oriented tree and the general signature}
\label{gentree}

\begin{figure}
\includegraphics[width=17cm]{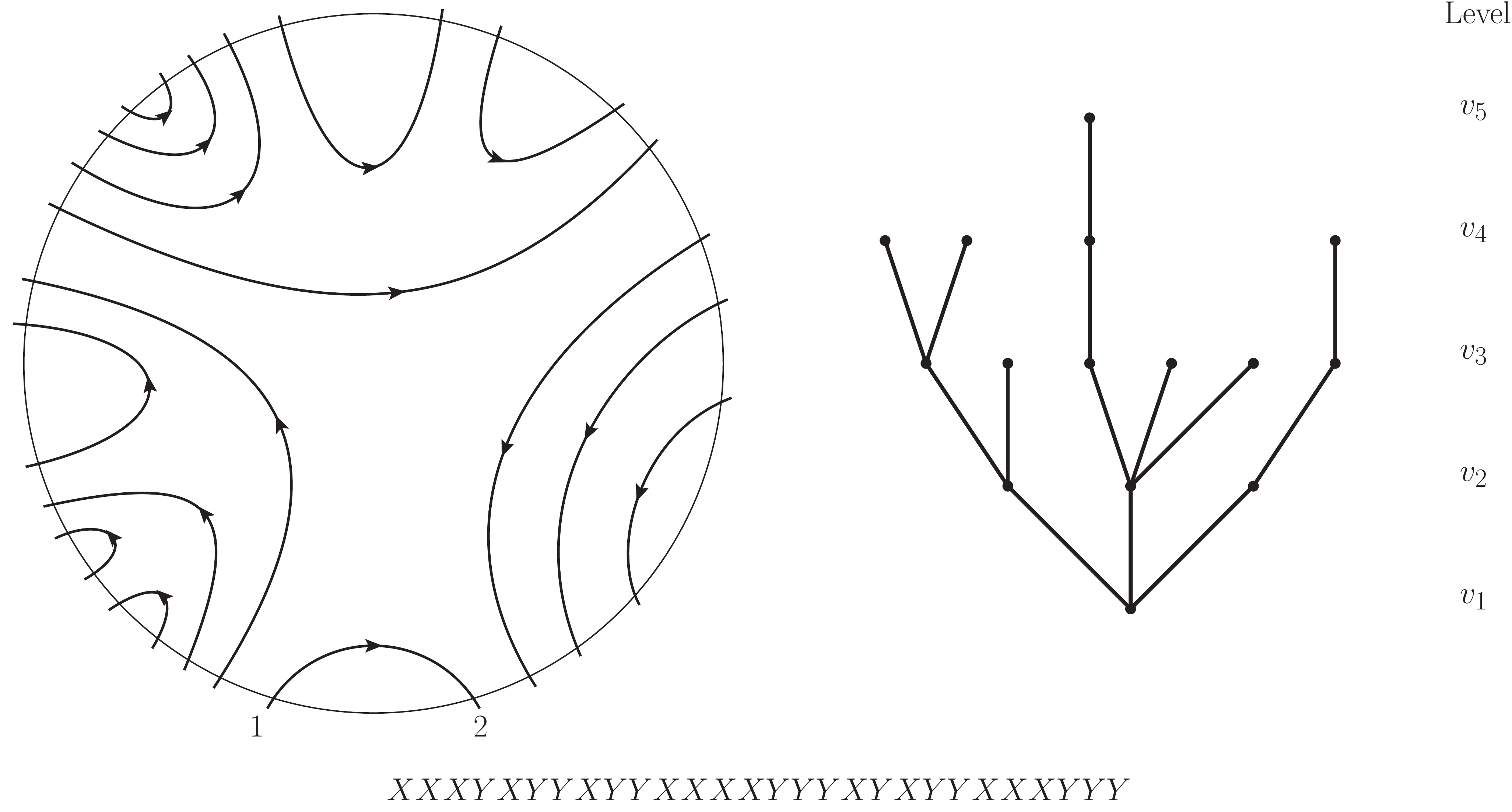}
\caption{The quark line graph for the above Dyck word is shown on the left. The quark line directions have been
chosen in accordance with an all-positive signature. On the right is the associated rooted oriented
Dyck tree, which can be seen as a dual graph to the quark line graph once the line $(1\to2)$ is removed and the circle representing
the edge of the plane is identified as a node. The levels
$v_i$ of the tree are given on the far right.}
\label{fig:tree}
\end{figure}

The second generalisation concerns the signature of the Dyck permutation --  which way the quark lines are directed 
in the quark line diagram for each primitive in the basis. With the prescription described here there
 are $2^{k-1}$ different bases; in \cite{Melia:2013bta}, only the basis
in which each permutation has all-positive signature $(+,+,\ldots,+)$ was proven. The more general result come from an
 iterative proof based around rooted oriented trees. These trees are also used in the organisation of the flavour recursion 
 in the following section. A rooted oriented tree is a dual graph to the quark line graphs with the edge $(1\to 2)$ removed, 
 with the circle identified as a node  to which the edges of the quark line graph (the quark lines) attach. 
 Alternatively, a rooted oriented tree can be drawn directly from the Dyck word via a `snail climbing up a tree' path -- 
 every time there is an $X$, the snail crawls up a branch, and every time there is a Y it crawls down the other side. 
 See Fig.~\ref{fig:tree}. The number of rooted oriented trees for a given number of nodes is given by the Catalan number, the same
 as the number of Dyck words.
 The rooted oriented tree is composed of nodes at different levels, labelled by $v_i$, where $1\le i\le k$, and with
 $i=k$ only being achieved for the Dyck word of the form $XX...XYY...Y$.
 
 An iterative procedure can be set up around the concept of the `maturity' of a tree, which is determined by the number of 
nodes $n_i$ the tree has at each level $v_i$, with a more mature tree having more nodes at higher levels. Concretely, we
can define a tree $A$ to be more mature than a tree $B$ if, comparing the number of nodes at each level $v_i$, starting at $i=1$,
some  $i$ is reached where $n_i^A<n_i^B$. It is possible that distinct trees can have the same maturity.

We can show that  the signature of a basis primitive can be chosen at will. For clarity of presentation we will make a couple of
simplifications. Firstly, we will present the proof for the pure-quark case, and point out the straightforward generalisations
to include gluons where necessary. Secondly, we will ignore the effect of Fermi statistics, which generates negative signs
for odd permutations of the labels from the canonical form $1\ldots n$ -- these signs can be put in by hand
after all relations are taken into account.
The main ingredient is being able to
show the following:
\be
\ma(1,\ldots,j,\beta,i,\ldots,2) =  -\, \ma(1,\ldots,i,\beta^T,j,\ldots,2)\,\, + \text{(trees of higher maturity)}\,,
\label{eqit}
\ee
where $\beta$ is some set of quark and antiquark labels, and where $\beta^T$
denotes the set $\beta$ with ordering reversed.
The `+(trees of higher maturity)' refers to primitives which have a more mature rooted oriented tree than the primitive on the lhs. 
The first primitive on the rhs has equal maturity to the primitive on the lhs -- the orientation of the quark pair $(i,j)$ has been
reversed, as has the orientation of all the quark pairs contained within $\beta$, but no other quark lines have had their
orientation changed. Eq.~\ref{eqit} follows from eq.~\ref{eq:app1} given in the appendix (with gluons this equation generalises
to eq.~\ref{eq:app2}), and
is a $SU(3)_c$ group theory relation -- it follows from  KK relations modified by the quark line structure.

\begin{figure}
\includegraphics[width=12cm]{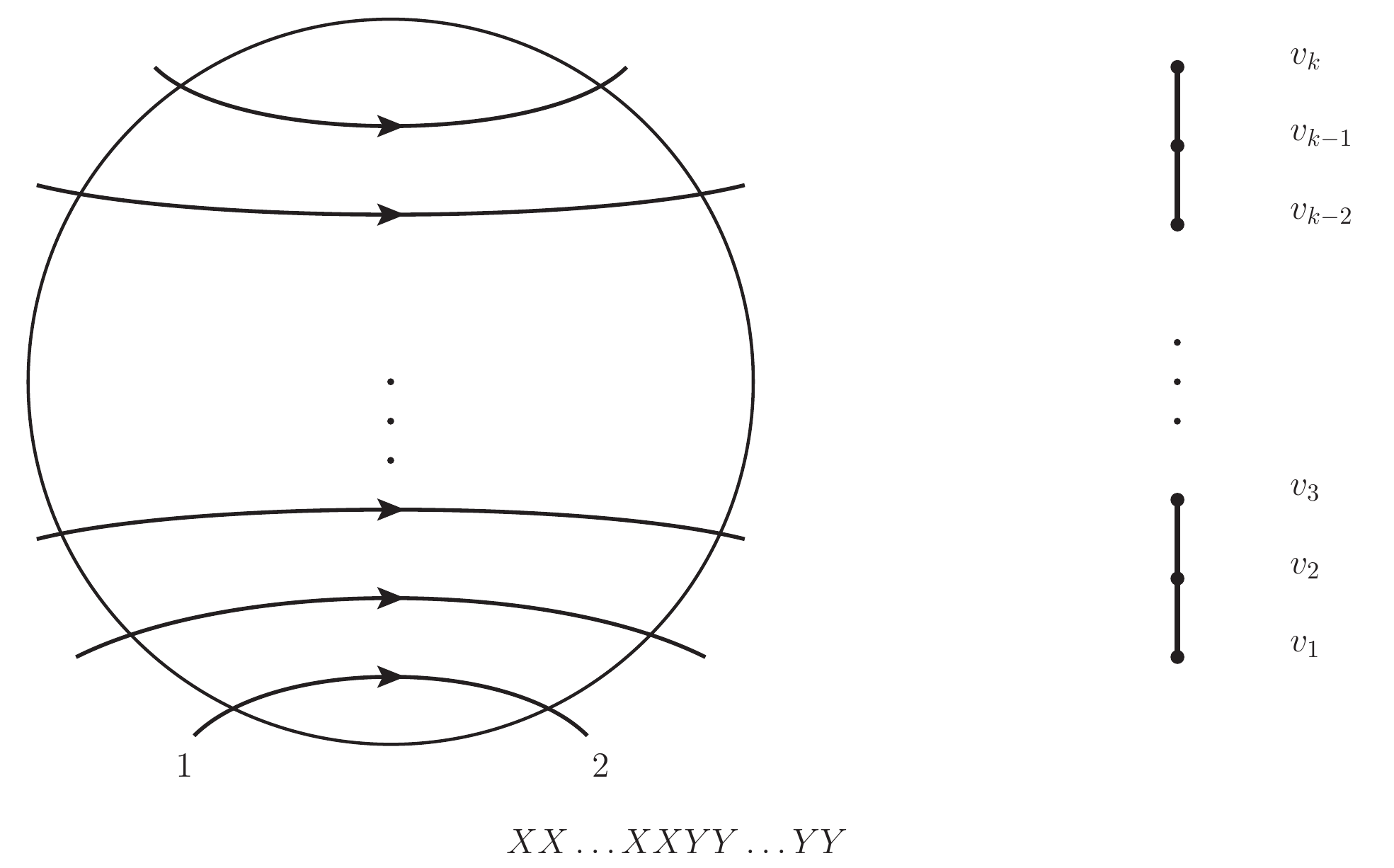}
\caption{The Dyck word and quark line graph of a primitive with $k$ quark lines, for
 which the rooted oriented Dyck tree has maximum height, or maturity.}
\label{fig:maxheight}
\end{figure}

If the quark lines of a primitive are oriented in a different way to the signature chosen for our basis 
vector for this particular Dyck topology and flavour assignment, we can apply eq.~\ref{eqit} to each of the wrongly
oriented flavour pairs, starting with the pair sitting at the lowest level on the rooted oriented tree (if there is more than
one pair at this level, they can each be oriented in turn, before orienting pairs further up the tree). We can then work upwards through
the tree, each time re-writing the primitive up to higher maturity terms, which we assume we can solve, until all of the
quark lines are oriented in the desired way, dictated by the signature of this particular flavour assignment that we chose for
our basis. This iterative procedure terminates at amplitudes which have a tree of highest possible maturity -- these
 are based on the Dyck word of the form $XX...XYY...Y$ (see Fig.~\ref{fig:maxheight}). These primitives can be ordered simply, since eq.~\ref{eq:app1} in this case becomes (see eq.~\ref{eq:app2} for the gluonic case)
 \be
 \ma(1,\ldots,j,\beta,i,\ldots,2) = - \, \ma(1,\ldots,i,\beta^T,j,\ldots,2)\,.
 \ee
 This completes the proof that the signature of each flavour pair assignment can be chosen independently.
As a simple example of how the iteration works, consider the pure six quark case, and chose as a basis from eqs.~\ref{eq1}-\ref{eq4}
where the first two have all-positive signature, and the second two have all-negative signature:
\be
\ma(1,3,4,5,6,2)\,,~~~\ma(1,5,6,3,4,2)\,,~~~\ma(1,6,4,3,5,2)\,,~~~\ma(1,4,6,5,3,2)\,.
\ee
We demonstrate how the primitive amplitude $\ma(1,4,3,6,5,2)$ can be expressed in terms of this basis.
Firstly, applying eq.~\ref{eq:app1},
\be
\ma(1,4,3,6,5,2) = -\ma(1,3,4,6,5,2) - \ma(1,3,6,5,4,2) \,.
\label{firstit}
\ee
Neither of the resulting terms are yet in our chosen basis. The iteration continues on the first of the primitives on the rhs of eq.~\ref{firstit} as
\be
\ma(1,3,4,6,5,2) &=& -\ma(1,3,4,5,6,2)-\ma(1,5,3,4,6,2) \nonumber \\ &=&  -\ma(1,3,4,5,6,2) - \bigg[ -\ma(1,6,4,3,5,2) \bigg] \,,
\ee
where in the second equality the recursion acts on the second primitive obtained after the first equality. The iteration
continues on the second of the primitives on the rhs of eq~\ref{firstit} as
\be
 \ma(1,3,6,5,4,2) = - \ma(1,4,5,6,3,2) = - \bigg[- \ma(1,4,6,5,3,2)\bigg] \,.
\ee
At this point the recursion has terminated, and we have expressed the primitive amplitude $\ma(1,4,3,5,6,2)$ in
terms of our basis:
\be
\ma(1,4,3,6,5,2) = \ma(1,3,4,5,6,2)-\ma(1,6,4,3,5,2) -\ma(1,4,6,5,3,2) \,.
\ee

The primitive amplitudes in the bases described in this section are independent, since we have shown how to express
any primitive in terms of a set of size $(n-2)!/k!$ -- this is the minimum size for a basis, since the one-flavour case with $(n-2)!$
independent primitives must be recovered using a  sum over the $k!$ momentum permutations of 
the distinct flavour case. That is, since
\be
\mathcal{M}_{\text{one-flavour}} = \sum_{\mathcal{P}(p_2,p_4,..,p_{2k})}(-1)^{\text{sgn}\mathcal{P}}\mathcal{M}_{\text{distinct-flavour}}(\{p_i\})\,,
\ee
the number of independent primitives $\ge (n-2)!/k!$.
The momentum permutations acting on the basis primitives of $\mathcal{M}_{\text{distinct-flavour}}$
bring them outside of the basis for a single distinct flavour amplitude. For instance the momentum swap $p_2\leftrightarrow p_4$
would mean that  the antiquark with momentum $p_1$ $(p_3)$ is no longer connected to the quark with momentum $p_2$ $(p_4)$.
This is different to swapping the cyclic positions of the quark with momentum $p_2$ and the quark with momentum $p_4$, but keeping
the same flavour pairing so that $p_1$ still connects to $p_2$ and $p_3$ still connects to $p_4$. The latter is what happens under
the KK relations, and was the subject of this section; different quark pairings, on the other hand, are tied up with what happens when quark lines have identical flavour,
and it is to these considerations we now turn.


\section{Flavour recursion and all massless QCD trees from $\sym$}
\label{flav}

The idea behind the flavour recursion is simple -- write a $k$-flavour primitive of the
form $\ma^{n_f=k}(1,\sigma,2)$ as a one-flavour primitive with the same labelling $\ma^{n_f=1}(1,\sigma,2)$,
and then subtract any wrong quark line contributions using $k$-flavour primitives with a different flavour pairing:
\be
\ma^{n_f=k}(1,\sigma,2)=\ma^{n_f=1}(1,\sigma,2) -\sum\ma_{ \text{wrong flav}}^{n_f=k}(1,\sigma,2)\,.
\label{eq:seventeen}
\ee
We can then iterate this procedure on each of the subtraction terms, but if 
this is to work, then these subtraction amplitudes must be further down an iterative direction which must
eventually terminate. It follows that there must exist some $k$-flavour primitive amplitudes that are simply equal to one-flavour primitives,
\be
\mak(1,\sigma,2)= \mao(1,\sigma,2)\,,
\label{eq:simpleplan}
\ee
and this happens for the permutations based 
around the Dyck tree of highest maturity (Fig.~\ref{fig:maxheight}) and which have an all-positive signature.
 We shall also see that the direction in which to iterate is up the rooted oriented tree introduced in the previous
section. 
The reason that such a relation as eq.~\ref{eq:simpleplan} exists is down to the role of planarity and cyclic ordering  
in restricting the pole structure of the one-flavour amplitude to be the same as that of the $k$-flavour.  Throughout this
section we use one-flavour to implicitly specify that all quark lines have the same helicity, which is the most
general case.

As a simple example of this planar-cyclic restriction, 
consider the four-quark primitive amplitudes $\mak(1,3,4,2)$ and $\mao(1,3,4,2)$. 
Although it is possible by flavour considerations
for a quark line to run from $(1\to4)$ in the latter primitive, this is forbidden by planarity and the specified
cyclic ordering:
\begin{center}
\includegraphics[height=3.5cm]{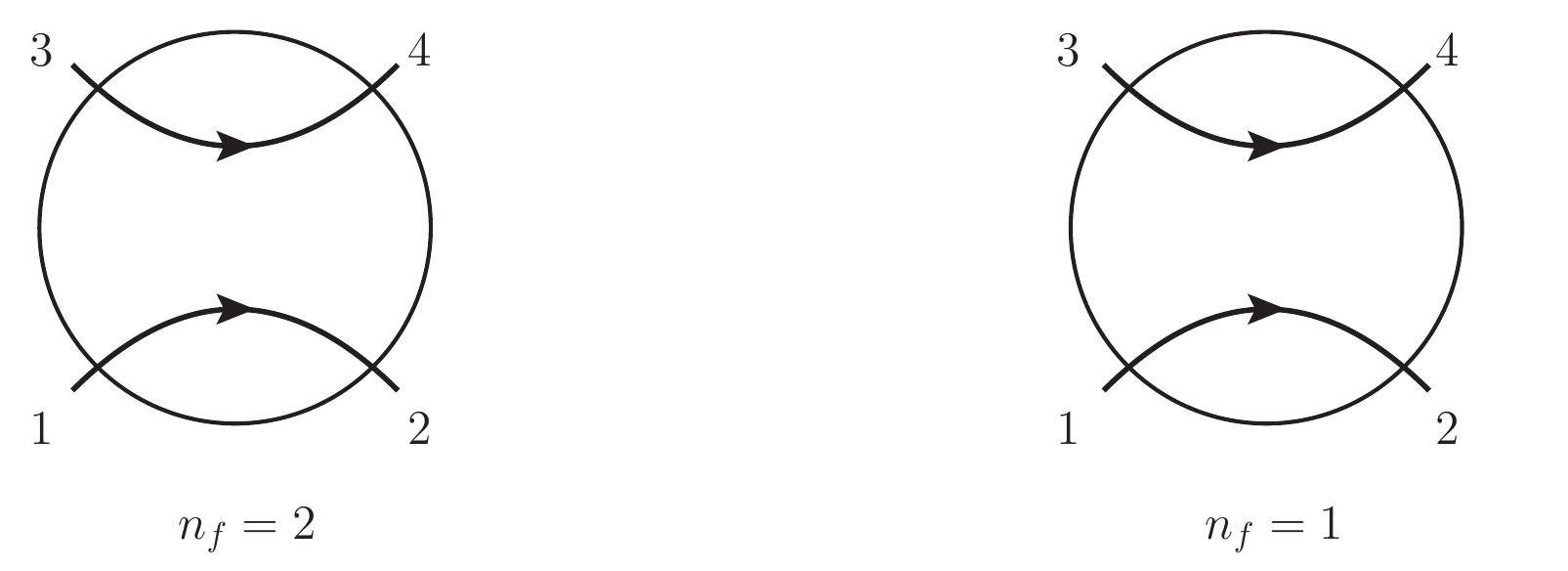}
\end{center} 
The situation is different if we compare the $k$-flavour amplitude $\mak(1,4,3,2)$ with
the one flavour amplitude with the same cyclic labelling, $\mao(1,4,3,2)$:
\begin{center}
\includegraphics[height=3.5cm]{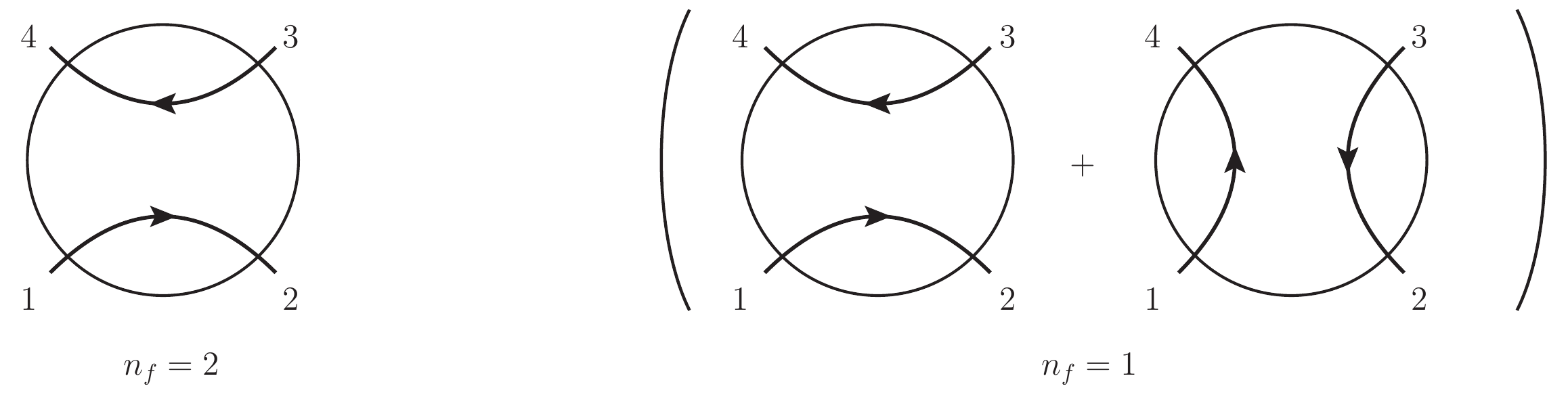}
\end{center}
Now there is a difference between the amplitudes -- in the one-flavour case there is both a $s$- and a $t$- channel pole, since nothing
restricts the quark line to run from $(1\to4)$. Again, for the sake of clarity, a simplification has been made to ignore
the minus signs arising from Fermi statistics when the  flavour pairing in the one-flavour amplitude permutes from $(1,2)(3,4)$
to $(1,4)(2,3)$ -- that is, a relative $+$ sign has been used between the quark line graphs of the one-flavour
primitive above, rather than a $-$ sign. Doing this will
make the structure of the flavour recursion in the next section more transparent --  the minus signs can easily be reinstated 
in whatever method is used to calculate the one-flavour amplitudes.
 
It is easy to see that planarity requires the one-flavour primitive amplitudes with a quark line structure based on $XX...XYY...Y$
 which have an all-positive signature to be identical to $k$-flavour primitives which have the same highest maturity tree:
\be
\mak(1 \,\bar{q}\,  \bar{q}\,\ldots  \bar{q} \,q\, q \ldots q\,2)=\mao(1 \,\bar{q}  \,\bar{q}\ldots  \bar{q} \,q \,q \ldots q\,2)\,.
\label{eq:recend}
\ee
There is no planar way in which to connect an antiquark  to a quark in the one-flavour case which is not the one it would
have connected to in the $k$-flavour case, without giving rise to crossed quark lines.

The all-positive signature basis will play a special role in the following. We now present the flavour recursion, which starts
by acting on a $k$-flavour primitives of all-positive signature:  
\begin{enumerate}
\item Express an all-positive signature $k$-flavour primitive as a one-flavour primitive which has the same cyclic ordering
of external particles, minus subtraction $k$-flavour primitives which serve to remove the wrong quark line contributions (see
eq.~\ref{eq:seventeen}).
\item Re-express each of the subtraction primitives in terms of all-positive signature $k$-flavour primitives as discussed in the
previous section. 
\item Repeat  from step 1. on each of these $k$-flavour primitives.
\end{enumerate}
Step 2 is important in order to allow the recursion to terminate.

As a warm-up to the general case for the recursion, 
consider again the purely six-quark primitive  $\mak_{(3,4)(5,6)}(1,3,4,5,6,2)$, which is an all-positive signature
 primitive. We have used the subscripts $(3,4)(5,6)$ to indicate the flavour pairing used (in the previous
 section this pairing would have been implicit). This primitive can be expressed
as a one-flavour primitive minus a subtraction as in the following way:
\vspace*{-0.6cm}
\begin{center}
\includegraphics[height=3.5cm]{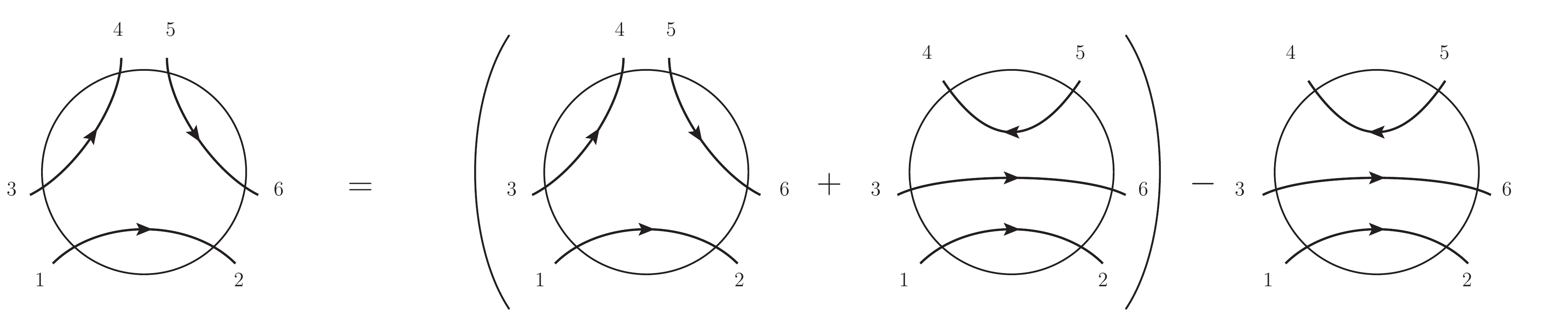}
\end{center}
\vspace*{-0.6cm}
\be
\mak_{(3,4)(5,6)}(1,3,4,5,6,2)=\mao(1,3,4,5,6,2) - \mak_{(3,6)(5,4)}(1,3,4,5,6,2) \,.
\label{6rec1}
\ee
Again, we are ignoring the effect of Fermi-statistics.
This completes step 1. We now need to re-express the subtraction primitive $\mak_{(3,6)(5,4)}(1,3,4,5,6,2)$ (which has a different flavour pairing as indicated
by the subscripts) in terms
of primitives of all-positive signature, as required by step 2 above.
This is achieved through the group theory relation eq.~\ref{eq:app1} (or with gluons eq.~\ref{eq:app2}),
\be
\mak_{(3,6)(5,4)}(1,3,4,5,6,2)=-\mak_{(3,6)(5,4)}(1,3,5,4,6,2)\,.
\label{6rec2a}
\ee
 The recursion now iterates (step 3) by acting with step 1 on this primitive:
\be
\mak_{(3,6)(5,4)}(1,3,5,4,6,2)=\mao(1,3,5,4,6,2)\,.
\label{6rec2}
\ee
 There are no further subtractions -- this is an amplitude with a Dyck tree of highest maturity -- 
 so the recursion terminates here and the full result is obtained, 
via eqs.~\ref{6rec2} and~\ref{6rec2a} into eq.~\ref{6rec1}:
\be
\mak_{(3,4)(5,6)}(1,3,4,5,6,2)=\mao(1,3,4,5,6,2)+\mao(1,3,5,4,6,2) \,.
\ee
Using this method, we have succeeded in expressing a three-flavour primitive in terms of one-flavour primitives.

As discussed at the end of the previous section, the intermediate primitive $\mak_{(3, 6)(5,4)}(1,3,5,4,6,2)$
is outside the basis of $k$-flavour primitives with flavour pairs $(3,4)(5,6)$, since it specifies 
that pairs of equal flavour are $(3\to6)$ and
$(4\to5)$.

\subsection{The flavour recursion for the general case}

The direction in which the general flavour recursion will iterate is based around the
rooted oriented tree introduced in Section~\ref{gentree}. 
We introduce a more streamlined notation to indicate the flavour structure of each primitive --
as in the above example, we will be dealing with $k$-flavour primitives with different flavour pairings
to the usual convention. We can label a primitive with a subscript $f=\{f_1,f_2,\ldots,f_k\}$ being a permutation
of $\{2,4,\ldots,2k\}$ to denote the flavour pairing $(1\to f_1), (3\to f_2),\ldots,(2k-1\to f_k)$. Under this notation, 
we can rewrite eq.~\ref{eq:seventeen} more precisely as
\be
\mak_{f}(1,\sigma, 2) = \mao(1,\sigma, 2) - \sum_{f' \in S_k} (1-\delta_{f f'})\,\mak_{f'}(1,\sigma,2) \,,
\label{firstsum}
\ee
where the $\delta_{ff'}$ removes the amplitude on the lhs from the sum on the rhs, and
 where we have again ignored minus signs coming from Fermi-statistics. In considering the full
 sum over $S_k$ we are taking into account all possible flavour pairings, but they will not all
 be non-zero, since some of these pairings will give rise to crossed quark lines. 

Now consider a general all-positive signature $k$-flavour primitive, for example, the one 
shown in Fig.~\ref{fig:tree}.
Which primitives in the sum on the rhs of eq.~\ref{firstsum} are zero? 
 When we are dealing with an all-positive signature permutation on the lhs,
 the reader can convince themselves that if the pairing $(1\to2)$ is not present, then the amplitude must be zero: 
 if the flavour pairing is $(1\to f_i)$ then there will be either an odd number of quarks or an odd number
 of antiquarks between the position $f_i$ and 1 in the cyclic order, $\mak(1,\ldots,f_i,\ldots,2)$. 
 
\begin{figure}
\includegraphics[width=10cm]{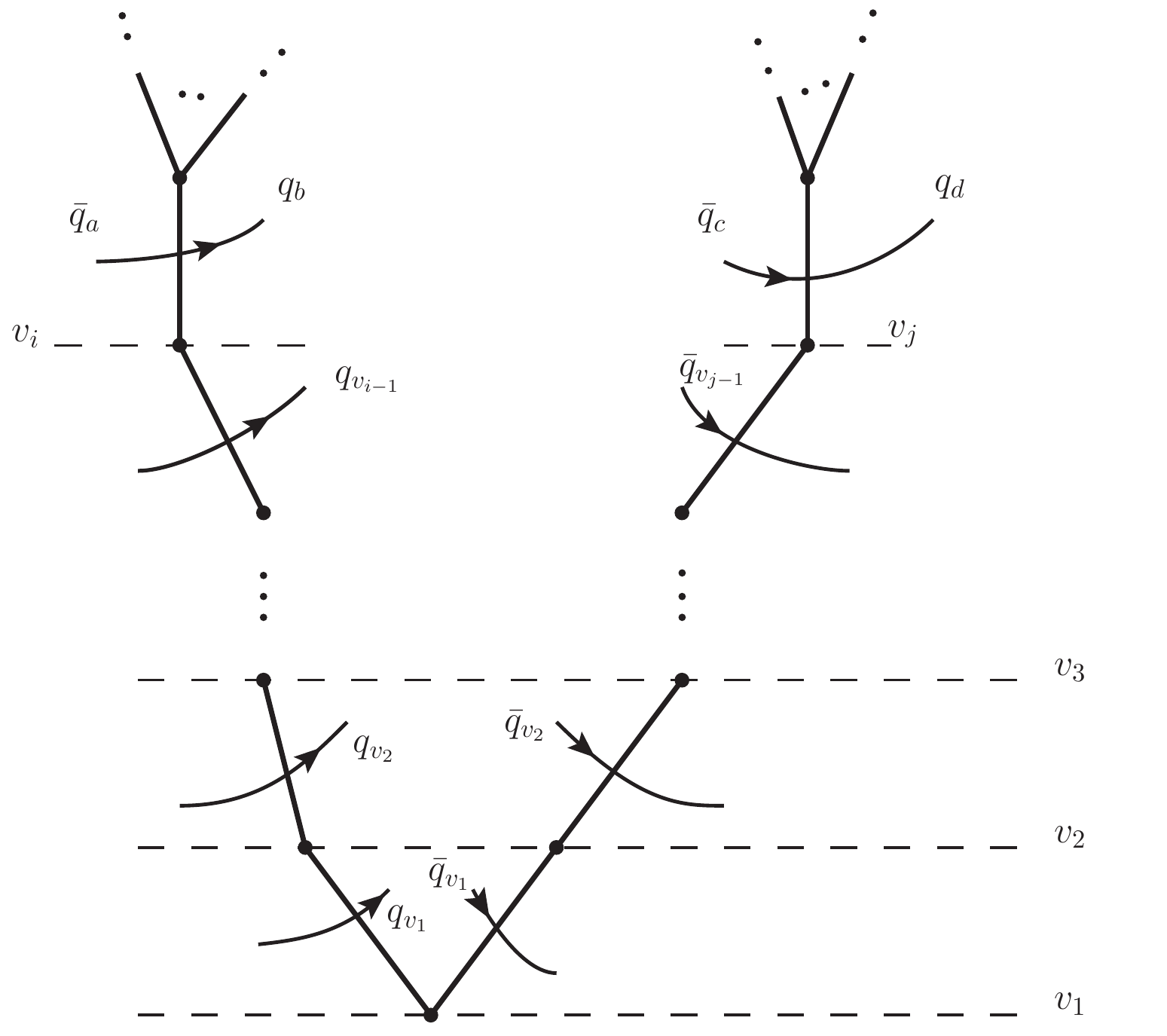}
\caption{Path back to the lowest node below two different nodes of the tree, one at level $v_i$, associated with quark line
$\bar{q}_1\to q_2$, and the other at level $v_j$, associated with the quark line $\bar{q}_3\to q_4$.
}
\label{fig:flavmix}
\end{figure}

 A more careful
 consideration reveals the following result: for the $k$-flavour subtraction primitives to be non-zero in eq.~\ref{firstsum} then
 the flavour permutations can only swap quark flavours at the same level of the Dyck tree. Consider any two nodes on the Dyck tree corresponding
 to the all-positive signature primitives on the lhs of eq.~\ref{firstsum}, one node at level $v_i$ and the other at level
 $v_j$. Next, consider the path to each node from to the root node at level $v_1$ (these paths are unique, but they could overlap up to some node $v^{\text{max}}_{\text{overlap}}$
 from which point onwards they differ -- in 
 this case in everything that follows, the node at $v_1$ should be taken to represent $v^{\text{max}}_{\text{overlap}}$)  -- see Fig.~\ref{fig:flavmix}. 
 Let $\bar{q}_a\to q_b$ denote the quark line which is the dual line passing through the edge above the node $v_i$, and 
 $\bar{q}_c\to q_d$ be the quark line which is the dual line passing through the edge above the node $v_j$. The primitive
 amplitude has the form 
 \be
 \ma_{ ..(a,b)..(c,d)..}(\ldots \bar{q}_a\ldots q_b \ldots \bar{q}_c \ldots q_d \ldots)\,.
 \ee
  The second set of ellipses 
 in this primitive contains quark and antiquark labels which are dual quark lines to edges in the tree above the node $v_i$. 
 They stand for an equal number of quarks and antiquarks. Similarly, the fourth set of ellipses in the primitive contain an
 equal number of quarks
 and antiquarks with quark lines dual to edges above the node $v_j$. Inbetween this, the third set of ellipses  could
 be any number of quarks and antiquarks coming from two possible origins. Firstly, they can come from lines
 dual to other edges of the Dyck tree which branch off to the right of the path to $v_i$ or lines dual to edges which branch
 off to the left of the path to $v_j$ -- this origin gives rise to an equal number of quarks and antiquarks. 
 Secondly they can be quarks 
 from the quark lines which cross the path to $v_i$ (labelled $q_{v_1},\ldots,q_{v_{i-1}}$ 
 in Fig.~\ref{fig:flavmix}) or they can be antiquarks from the quark lines which cross the 
 path to $v_j$  (labelled $\bar{q}_{v_1},\ldots,\bar{q}_{v_{j-1}}$).
 The fact that it is always quarks crossing the path to $v_i$ whereas it is antiquarks crossing the path to $v_j$ is a consequence
 of having a primitive with an all-positive signature -- in this way it is an important feature of the recursion. There are $i-1$ quarks
 and $j-1$ antiquarks from this second origin.
 
 Now consider what happens when the quark lines $\bar{q}_a\to q_b$ and  $\bar{q}_c\to q_d$ mix, so that
 we have $\bar{q}_a\to q_d$ and $\bar{q}_c\to q_b$. The number of quarks and antiquarks in-between $\bar{q}_a$ and $q_d$
 needs to be equal for the primitive $\ma_{..(a,d)..(c,b)..}(\ldots \bar{q}_a\ldots q_b \ldots \bar{q}_c \ldots q_d \ldots)$ to
 be non-zero, so it is a necessary condition that $v_i=v_j$.

 We can use this information to split up the sum over all mixings in eq.~\ref{firstsum} as follows. Separate the labelling of
 the flavour structure into lists of those at equal levels of the tree: $f_1\ldots f_h$, where $h$ is the highest level of the tree,
 so that the primitive can be labelled $\mak_{f_1\ldots f_{h}}$.
We can then write the permutation sum by splitting it up so that only permutations within each level are considered
 \be
 \mak_{f_1f_2..f_h}(1,\sigma,2)=\mao(1,\sigma,2)-\sum_{f'_1\in S_{n_1}}\sum_{f'_2\in S_{n_2}}\ldots\sum_{f'_h\in S_{n_h}} (1-\prod_{i=1}^h\delta_{f_if_i'}) \,\,\mak_{f'_1f'_2..f'_h}(1,\sigma,2) \,.\nonumber\\
\label{thefeq}
 \ee
  Again, the product of delta functions serves to remove 
 the amplitude on the lhs from the sum. It is just the quark line structure which is important for the recursion; no reference to the gluon position is needed.
In the sum on the rhs some of the amplitudes are still zero, but splitting the sum up explicitly as above  makes it clear that a quark at a higher level of the tree cannot pair with an antiquark at a lower level of the tree,
and as such all of the $k$-flavour amplitudes appearing on the 
rhs of eq.~\ref{thefeq}  have a Dyck tree of higher maturity than the primitive on the lhs. Again, this feature is a consequence of having of
 having an  all-positive signature primitive on the lhs.


Eq.~\ref{thefeq} is the master equation for the recursion. After it is applied,
the recursion continues by next expressing each of the subtracted $k$-flavour amplitudes on the rhs of eq.~\ref{thefeq} in their 
all-positive signature Dyck basis. As discussed in the previous section, in doing this the only primitives of higher maturity
are created. This completes one iteration of the recursion. After it, all remaining $k$-flavour 
amplitudes are of higher maturity than the one of the lhs of eq.~\ref{thefeq}. The recursion eventually terminates at the amplitude 
of maximum height, Fig.~\ref{fig:maxheight}, through eq.~\ref{eq:recend}.

\subsection{Obtaining QCD from $\sym$}
\label{sym}

\begin{figure}
\includegraphics[width=5cm]{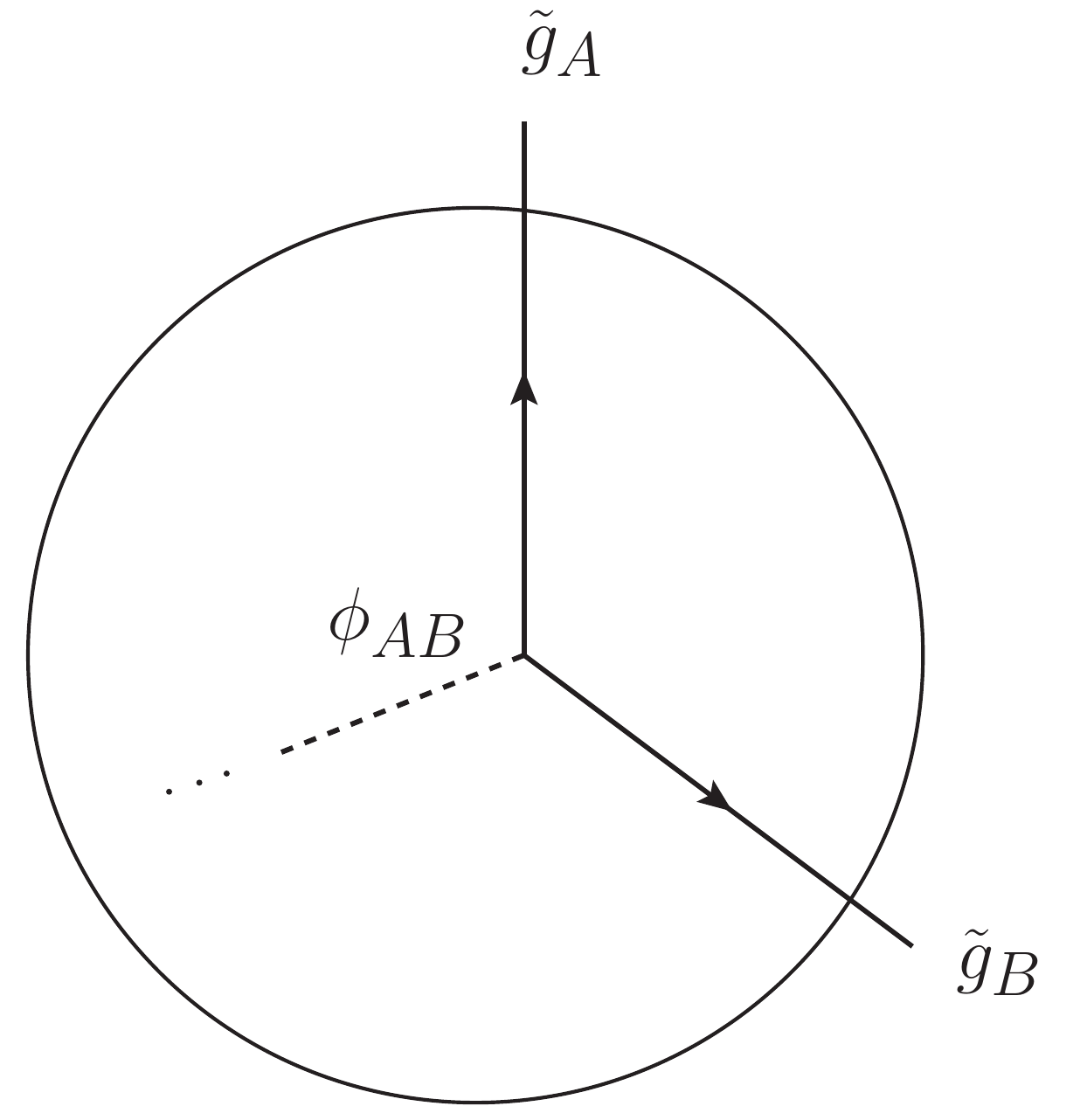}
\caption{The termination of a scalar $\phi_{AB}$ results in two external gluinos $\tilde{g}_A$ and  $\tilde{g}_B$ (to be identified as quarks)
 of different flavours, since the coupling requires $A\ne B$.}
\label{fig:sym}
\end{figure}

We do not present the details of Dummond and Henn's formula for the solution to $\sym$ at tree-level, 
and instead refer the reader to the original publication, \cite{Drummond:2008cr}. 
The solution is given in terms of a super wave 
function $\Phi$, which in terms of on-shell gluon ($g^+$, $g^-$), 
gluino ($\tilde{g}_A$, $\bar{\tilde{g}}^A$), and scalar ($\phi_{AB}$) states, and Grassmann variables $\eta^A$, is
\be
\Phi(\lambda,\bar{\lambda},\eta)= &g^+\lcl+ \eta^A \tilde{g}_A\lcl + \frac{1}{2}\eta^A\eta^B\phi_{AB}\lcl ~~~~~~~~~~~~~~~~~\nonumber \\
&+ \eta^A  \eta^B   \eta^C \epsilon_{ABCD} \, \bar{\tilde{g}}^D\lcl +\eta^A  \eta^B   \eta^C  \eta^D \epsilon_{ABCD} \,g^-\lcl
\ee
with $A=1,2,3,4$. In \cite{Drummond:2008cr} it was discussed how to perform the Grassmann integrations so as to 
project the formula onto specific external states, and in  \cite{Dixon:2010ik}, all-$n$ formulas were derived for the projection
onto $n$ external particles -- gluons and gluinos
of the four possible flavours, $A=1, 2, 3, 4$. These amplitudes then 
begin to resemble QCD amplitudes, with gluinos identified as quarks -- since they
are colour stripped primitives, the gluinos have identical interactions with gluons as do quarks. The problem in relating these
amplitudes to the QCD amplitudes lies in avoiding non-QCD interactions involving scalars, which couple two gluinos of different
flavour -- see Fig.~\ref{fig:sym}. This was achieved for specific cases in \cite{Dixon:2010ik} by careful choices of external flavour
 and by summing over different flavour permutations so that scalar contributions were eliminated -- either not being present or cancelling 
 against each other -- so that all QCD 
 amplitudes with up to four distinct flavour quark lines
 were shown to be obtainable from Drummond and Henn's solution. It was left as an open problem as to whether amplitudes
 with more quarks could also be obtained -- clearly a necessary issue to deal with is the one of flavour, since there are only four
 flavours of gluino and if the $\sym$ solution were able to describe five-flavour amplitudes then these would have to be obtainable
 from at least four-flavour amplitudes.

It follows directly from the flavour recursion described in the previous section that in fact the whole of 
massless QCD at tree-level {\it is} obtainable from $\sym$, since
one-flavour amplitudes are identical in QCD and $\sym$. This is because if a scalar exchange is created, we can trace it
to its termination point at some quark line (it has to terminate, because no external scalars are specified in the amplitude), 
see Fig.~\ref{fig:sym}. However, for this to be non-zero, it necessarily creates two different flavoured 
quarks which will eventually leave the amplitude, contradicting the original specification of a one-flavour amplitude. 
In other words, when all
quark lines are of the same flavour, no scalars can be exchanged between the quark lines, since they couple to different
flavoured quarks as $\phi_{AB} \,\,\tilde{g}_A\, \,\tilde{g}_B,\,A\ne B$. (This is a similar reasoning to that which
asserts that all-gluon tree amplitudes
are identical in QCD and $\sym$).

So, in order to obtain all of massless QCD at tree-level, the most difficult primitives to obtain, the all-distinct flavour
cases, are expressed through the flavour recursion in terms of a particular set of one-flavour QCD primitives. These can
then be directly obtained from the one-flavour $\sym$
amplitudes, once projected onto gluon and one-flavour gluino external states, as described in  \cite{Dixon:2010ik}. 
One feature of this particular set of one-flavour primitives is that  the external quark and antiquark labels are
fixed into a Dyck word, with $X$s identified with antiquark labels and $Y$s identified with quark labels.

It is also possible to set up a second recursion from which to obtain QCD trees from $\sym$ trees. This is a two-flavour
recursion, and is described briefly in
appendix \ref{append2}.

\section{Discussion}
\label{disc}

Before concluding,  we address issues surrounding 
amplitudes beyond massless QCD, as well as the role of
 BCJ relations between the one-flavour primitives, and the counting of
 the number of one-flavour primitives needed to produce the $k$-flavour primitives.

\subsection{Beyond massless QCD}
\label{otherps}

In this subsection we describe the behaviour of amplitudes involving electroweak bosons
and massive quarks specifically under the KK relations and the flavour recursion. 

Amplitudes involving $W$s, $Z$s and photons all require external quark lines in tree-level
amplitudes to couple to. 
The electroweak particles do not carry colour charge, and these amplitudes satisfy the KK relations 
described in previous sections as if there were no EW particle present. When there is more than one 
weak boson present, the amplitudes can be further ordered under their $SU(2)$ charge -- see Ref.~\cite{Dai:2012jh}.
However, because quarks of different flavour carry different electroweak charge, the flavour
recursion cannot apply once electroweak bosons are present -- they are sensitive to flavour in contrast to gluons which are not.

Amplitudes involving massive quarks, top quarks for example, 
also satisfy the same KK relations as those with purely massless quarks -- the presence of a massive quark 
line does not affect the group theory factors from which the relations follow. 
However, here too the flavour recursion relation described above
will not work, since a one-flavour amplitude with external quark lines of differing mass is not well defined.

\subsection{BCJ relations and QCD primitives}
\begin{table}
  \begin{tabular}{|c|c|c|c|}
  \hline
   $n,k$ & $\#$ $k$-flavour (no BCJ) & $\#$ one-flavour for $k$-flavour (no BCJ) & BCJ one-flavour\\
   \hline 
   4,2 & 1 & 1 & 1 \\
   6,3 & 4 & 6 & 6 \\
   8,4 & 30 & 90 & 120 \\
   10,5 & 336 & 2520 & 5040 \\

...    & ... & ... & ... \\

      $n,k$ & $(n-2)!/k!$ & $(n-2)!/2^{k-1}$ & $(n-3)!$ \\
\hline
  \end{tabular}
   \caption{Counting the number of independent primitive amplitudes under KK relations for the distinct flavour case, the number
   of one-flavour amplitudes needed to construct a basis of these amplitudes, and the number of independent primitive amplitudes under BCJ
   relations for the one-flavour case.}
 \label{tbl:1}
\end{table}

Another interesting thing about being able to express all of massless QCD in terms of
one-flavour amplitudes, is that that these amplitudes can be written
in BCJ form, with kinematic numerator factors satisfying Jacobi relations such that the number
of independent primitive amplitudes is reduced to $(n-3)!$ (relations hold between primitives
multiplied by kinematic factors).  

This motivates a question as to how BCJ relations apply in the distinct flavour case. This
would have to be reconciled with the counting of the number of one-flavour primitives needed
to reconstruct a basis for the distinct flavour case. We perform this counting in Table~\ref{tbl:1}, where this number is compared
to the number of $k$-flavour independent primitives under KK relations alone, and the $(n-3)!$ one-flavour primitives under BCJ. 
The second column was constructed using examples up to the pure quark case $n=10$ -- i.e. using the flavour recursion to
find the one-flavour primitives needed for every $k$-flavour amplitude and then taking the union of this set 
 -- and the general formula is conjectured to hold beyond this. 
 
 Are there BCJ relations between the primitives in this particular set of all the one-flavour primitives? 
 If so, how many independent amplitudes remain? These questions will be interesting to address in future work.

\subsection{Further discussion}

We have shown that any tree-level massless QCD amplitude can be expressed
in terms of amplitudes that possess effective $\mathcal{N}=1$ supersymmetry. 
These amplitudes might be expected to have nicer properties than the distinct flavour
case, owing to this fact. Of course, one such realisation of this is that a closed
form solution for them is known, this in turn owing to the fact that $\mathcal{N}=1$ SYM
is a closed subset of $\sym$, for which this tree-level solution is known. 

The flavour recursion singles out a particular set of one-flavour amplitudes 
needed to reconstruct the distinct-flavour amplitudes. These are the ones with
the external quarks and antiquarks ordered as Dyck words, with antiquarks 
at $X$ locations and quarks at $Y$ locations. One might wonder whether
these amplitudes could take a simpler form, given that the underlying 
Feynman diagram representation for $k$-flavour amplitudes is considerably 
simpler than the one-flavour case. Interestingly, in Drummond
and Henn's proof of their solution to $\sym$ trees, the concept of a rooted (but not oriented) 
tree was also used to define
a direction in which to perform BCFW recursion, and there is some freedom
in this choice of recursive direction. It would be interesting if some alignment of the BCFW recursion 
with the Dyck tree could result in expressions where the quark line structure was
easily identified in the analytic form. Another avenue would be to investigate the use of the momentum twistor 
variables \cite{Hodges:2009hk,Drummond:2010qh} $(\lambda,\mu,\chi)$ rather than the $(\lambda,\tilde{\lambda},\eta)$ of Ref.~\cite{Drummond:2008cr}, 
under which 
expressions for amplitudes involving fermions can take simpler form.

A further feature of the Kleiss-Kuijf relations worth mentioning 
is that they do not depend on the dimensionality of spacetime. Studies of amplitudes in different spacetime dimensions 
is of theoretical interest, in particular in $d=3$ and $d=6$ (see e.g. \cite{Elvang:2013cua} for a recent review),
and dimensional regularisation is a common way of isolating infra-red and ultraviolet singularities in amplitudes. 
For phenomenological applications, the technique of 
$D$-dimensional generalised unitary \cite{Giele:2008ve} is an approach to build one-loop amplitudes out of tree-level primitive
amplitudes in dimensions $d>4$, and is particularly suited to  numerical evaluation.

 Since then the flavour recursion also does not depend on the dimensions
 of spacetime, and given that unitarity based techniques are able
to construct any loop-order amplitude purely out out of trees (evaluated, in general, in higher dimensions), 
 it follows that any loop-order amplitude in massless QCD is obtainable from tree-level amplitudes
  calculated in one-flavour QCD.
 This  means that the only objects that actually require a field theory calculation (i.e. the tree-level amplitudes)
 are done in a theory where there is no notion of flavour. Rather, flavour
  comes about only in the way we combine these objects, as we have seen
  for the tree-level case studied  in this paper.

\section{Conclusion}
\label{conc}

In conclusion, we have studied the  $SU(3)_c$ group theory relations between  tree-level QCD
primitive amplitudes  involving $k$ quark lines of distinct flavour and $n-2k$ gluons, and shown that they reduce
the number of independent primitives to $(n-2)!/k!$. We described how bases can be constructed  using the concept 
of a rooted, oriented Dyck tree. Exploiting the planarity and cyclic ordering of a class of these primitives which
have an all-positive signature, and using the $SU(3)_c$ relations,
 we derived a flavour recursion relation, also based around a Dyck tree. This flavour recursion can express
 a  multi-flavour tree-level QCD primitive in terms of one-flavour tree-level QCD primitives,
 which possess effective $\mathcal{N}=1$ supersymmetry. In turn, this makes it possible to use known formulas
 from $\sym$  to obtain all of massless QCD at tree-level. An interesting aspect is that using these results, no notion
 of flavour is needed when making amplitude  calculations in perturbative, massless QCD, given that unitarity based techniques
 can construct higher loop amplitudes out of tree-level amplitudes. 

We expect that a knowledge of a minimal general QCD basis, along with implementation of one-flavour $\sym$ results,
will increase the performance of computational efforts to describe multi-jet events at fixed order in
perturbation theory (see \cite{Badger:2012uz}
for a discussion of the phenomenological use of the $\sym$ analytic formulas, although the use of purely one-flavour formulas 
should be investigated further). 
A mathematica package implementing 
 for expressing any QCD primitive in terms of a chosen basis
of size $(n-2)!/k!$, and for implementing the flavour recursion relation to express a general primitive amplitude
 in terms of one-flavour amplitudes, will be made available on the website \cite{myweb}.
\newline

\noindent In the final stages of the preparation of this script, the preprint \cite{Schuster:2013aya} appeared, which
  provides an independent, multi-flavour approach to obtaining QCD trees from $\sym$, similar to the two-flavour 
  one outlined in appendix~\ref{append2} of  this work. 
  It would be interesting to fully explore this connection, and the one between the fermion flip identities defined in \cite{Schuster:2013aya}
  and the $SU(3)_c$ relations presented in Ref.~\cite{Melia:2013bta} and used here.

\begin{acknowledgments}
This research was supported by ERC grant 291377 ``LHCtheory -- Theoretical predictions and analyses of LHC 
physics: advancing the precision frontier''. I would like to thank Mat Bullimore, Lance Dixon, James Drummond,  
Johannes Henn, and Henrik Johansson for interesting and helpful conversations and suggestions during the course of this work.
\end{acknowledgments}

\appendix

\section{Group theory identities between QCD primitives}
\label{append1}

A group theory identity for the pure multi-quark case, $2k=n$, which can be used to further relate 
primitive amplitudes of the form $\ma(1,\ldots,2)$ was given in \cite{Melia:2013bta}, where a proof is also given -- see this
reference for a detailed discussion. The identity is given with reference to a primitive of the form
\vspace*{-0.3cm}
\begin{center}
\includegraphics[width=15cm]{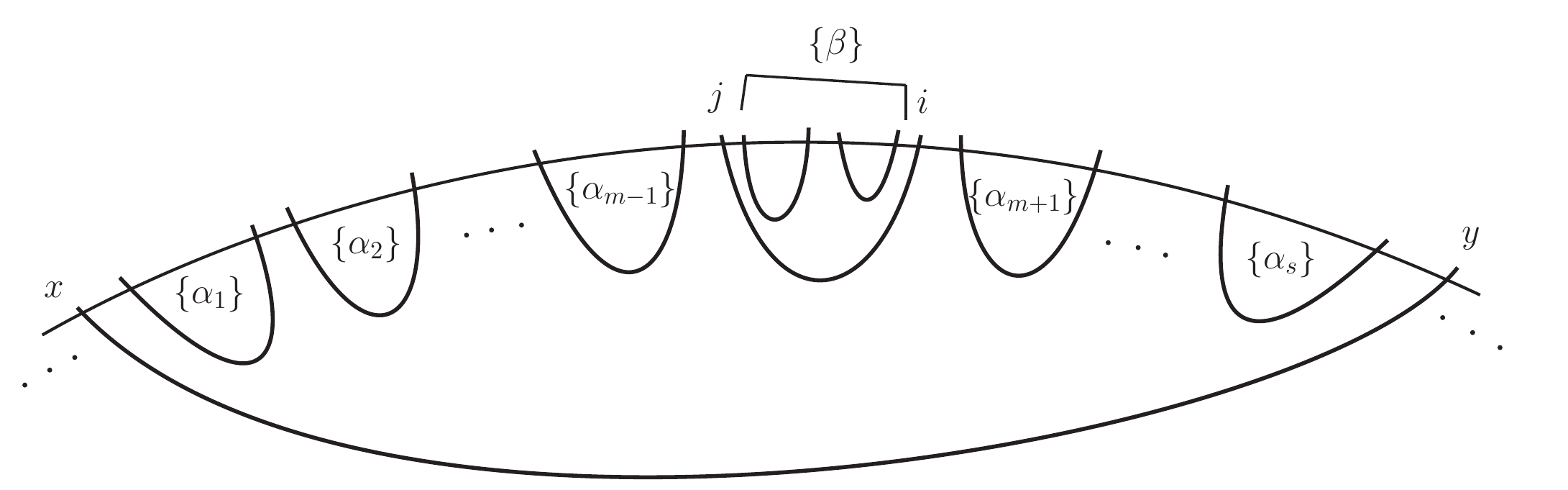}
\end{center}
\vspace*{-1.2cm}
\be
\ma(\ldots\, x\, \{\alpha_1 \} \,\{\alpha_2 \} \, \ldots \, \{\alpha_{m-1} \} \,j \, \{\beta\}\, i \,  \{\alpha_{m+1} \} \,\ldots\,  \{\alpha_{s} \} \,y \,\ldots)\,.
\ee
Here, the boundaries of the sets $\{\alpha_i\}$ are quark lines corresponding to a level one higher than the level of
 the quark line $(x-y)$ in the Dyck tree for the above primitive. Each set $\{\alpha_i\}$ may contain further quark lines (i.e. they
 may be sets of length greater than two). The quark line $(j-i)$ is also at a level one higher in the Dyck tree than the quark line $(x-y)$.
 There can be further quark lines between the positions of the labels $j$ and $i$ in the primitive -- these are denoted 
 as $\{\beta\}$ (they are all at a higher level in the Dyck tree than $( j-i)\,$). A group theory identity satisfied by this primitive is
\be
\ma(\ldots x\,\{\alpha_1\}..\{\alpha_{m-1}\}\,j\,\{\beta\}\,i \,\{\alpha_{m+1}\}..\{\alpha_{s}\} \, y\ldots) = ~~~~~~~~~~~~~~~~~~~~~~~~~~~~~~~~~~~~~~~~~~~~~~~~~~~~~~~~~~~~~ \nonumber\\
-\sum_{c=1}^{m}\bigg[ \sum_{\text{OP}\{D_c\}\{E\}}\bigg(\sum_{\text{OP}\{A_c\}\{B\}}\ma(\ldots x\,\{\alpha_1\}..\{\alpha_{c-1}\}\,i\,\overbrace{\underbrace{\{\alpha_{c}\}..\{\alpha_{m-1}\}}_{\{A_c\}}\underbrace{\{\beta^T\}}_{\{B\}}\,j}^{\{D_c\}} \,\overbrace{\{\alpha_{m+1}\}..\{\alpha_{s}\}}^{\{E\}}  \,y\ldots) \bigg)\bigg]\,, \nonumber \\
\label{eq:app1}
\ee
where $\text{OP}\{A \}\{B \}$ stands for `ordered permutations' and is the shuffle product of the sets $\{A\}$ and $\{B\}$, and where
$\{\beta^T\}$ is the set $\{\beta\}$ with the ordering reversed. In all of the primitives on the rhs, the direction of the quark
line $(j=i)$ has been reversed. Not all of the primitives generated by the ordered permutations are non-zero -- the
shuffle product can give rise to some configurations with crossed quark lines.
As with the expressions in the main body of this paper, eq.~\ref{eq:app1} should be dressed with appropriate 
minus signs as dictated by Fermi statistics.

If gluons are also present, the identity generalises as follows. Let ${\bf g_1}$ be the set of gluons inbetween $x$ and the set $\{\alpha_1\}$,
 ${\bf g_2}$ be the set of gluons in-between the set $\{\alpha_1\}$ and $\{\alpha_2\}$, and so on. The most general primitive amplitude 
 takes the form 
 \be
 \ma(\ldots x\, {\bf g_1}\, \{\alpha_1\} \,  {\bf g_2}\, \{\alpha_2\} \, .. \, {\bf g_{m-1}}\,\{\alpha_{m-1}\}\, {\bf g_m}\,j\,\{\beta\}\,i \,{\bf g_{m+1}}\,\{\alpha_{m+1}\}\,..\,{\bf g_s}\,\{\alpha_{s}\} \, {\bf g_{s+1}}\,y\ldots)\,.
 \ee
 The first and last elements of the sets $\{\alpha_i\}$ are by construction
quark labels, but we now allow for the possibility that any other element of the sets $\{\alpha_i\}$ 
as well as any element of the set $\{\beta\}$ 
to be a gluon label. Denote the first $i$ elements in the set ${\bf g_c}$ as
the set ${\bf g_c^i}$, and the remaining $n_c-i$ elements as ${\bf g_c^{n_c-i}}$, where $n_c$ is the number of gluons
in the set ${\bf g_c}$.  The generalisation of eq.~\ref{eq:app1} is
\be
&~&\ma(\ldots x\, {\bf g_1}\, \{\alpha_1\} \,  {\bf g_2}\, \{\alpha_2\} \, .. \, {\bf g_{m-1}}\,\{\alpha_{m-1}\}\, {\bf g_m}\,j\,\{\beta\}\,i \,{\bf g_{m+1}}\,\{\alpha_{m+1}\}\,..\,{\bf g_s}\,\{\alpha_{s}\} \, {\bf g_{s+1}}\,y\ldots) = \nonumber\\
&~&(-1)^{1+n_{g_\beta}}\sum_{c=1}^{m}\bigg[ \sum_{i=0}^{n_{g_c}}\bigg[ \sum_{\text{OP}\{D^i_c\}\{E\}}\bigg(\sum_{\text{OP}\{A^i_c\}\{B\}} \nonumber \\ 
&~&\ma(\ldots x\,{\bf g_1}\, \{\alpha_1\}\,..\,{\bf g_{c-1}}\, \{\alpha_{c-1}\}\,{\bf g^{i}_c}\, i\,\overbrace{\underbrace{{\bf g^{n_c-i}_c}\, \{\alpha_{c}\}\,..\,{\bf g_{m-1}}\, \{\alpha_{m-1}\}\,{\bf g_{m}}  }_{\{A^i_c\}}\underbrace{\{\beta^T\}}_{\{B\}}\,j}^{\{D^i_c\}} \,\overbrace{{\bf g_{m+1}}\, \{\alpha_{m+1}\}..{\bf g_s}\, \{\alpha_{s}\}{\bf g_{s+1}}\, }^{\{E\}}  \,y\ldots) \,\,\,\, \bigg)\bigg]\bigg]\,,   \nonumber \\
\label{eq:app2}
\ee
where $n_{g_\beta}$ is the number of gluon labels contained in the set $\{\beta\}$.
The proof of this equation is a straightforward generalisation of the proof of eq.~\ref{eq:app1}, given in \cite{Melia:2013bta}. The quark line structure generated in eq.~\ref{eq:app2} is exactly
the same as the quark line structure generated in eq.~\ref{eq:app1}, the only difference being there are more terms involving
each quark line structure,  with different placings of gluons which now enter the shuffle. 
As in eq.~\ref{eq:app1}, some of the terms generated by the shuffle products will be zero because of quark line constraints.

\section{A two-flavour recursion to obtain QCD from $\sym$}
\label{append2}

\begin{figure}
\includegraphics[width=15cm]{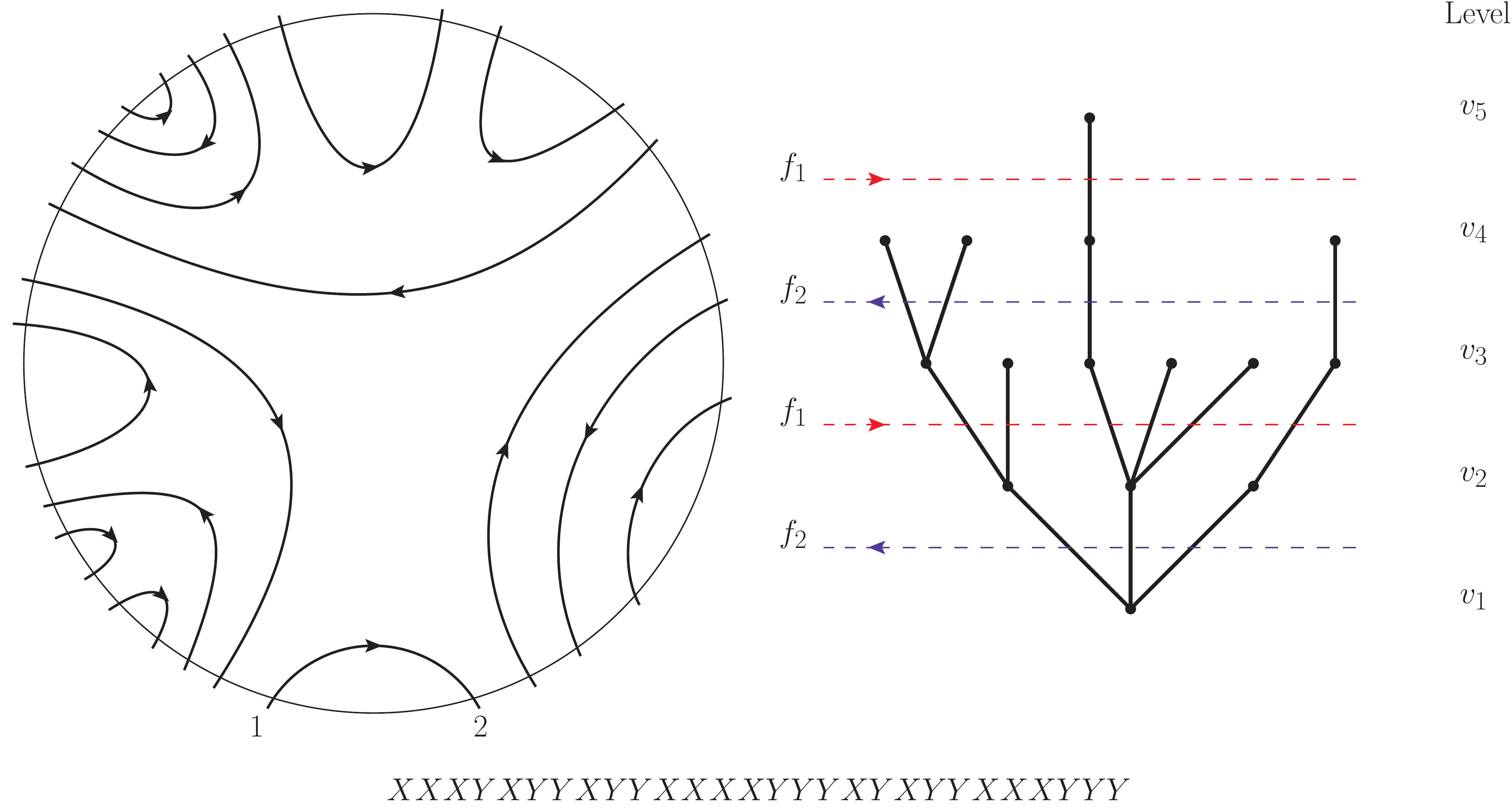}
\caption{The quark line graph on the lhs shows how to choose the signature of the Dyck permutation 
(i.e. quark line direction) so as to create alternating quark and antiquark external states. The rhs is the corresponding
Dyck tree. The dotted lines drawn across the edges of the Dyck tree have an arrow to indicate the direction of the arrows
on the quark lines which cross them to create the quark line graph shown on the left, and have a flavour label $f_1$
or $f_2$ to indicate which flavour the quark lines should have in the two-flavour recursion described in the text.}
\label{fig:twosym}
\end{figure}

In this appendix we briefly outline a flavour recursion based on a different basis to the all-positive 
signature recursion of Sec.~\ref{flav}. The choice of signature for each Dyck topology is the one which
results in alternating quarks and antiquarks for the external particles. In this way
the signature of each primitive will depend on its Dyck topology. An example graph is shown in Fig.~\ref{fig:twosym}. The
reason for choosing such a basis is that it is another way to prevent any possible scalar exchange between gluino lines
in the $\sym$ amplitudes. Consider following an internal
scalar line up until the point at which it terminates, as shown in Fig.~\ref{fig:sym}. Now we can relax the requirement
that all external quarks (we use quark in this section, but this should be identified with gluino in the $\sym$ context) 
are the same flavour, so that such a scalar contribution to the primitive could exist. However,
if the alternating basis is chosen, then since the termination point is two quarks (or two antiquarks) there
will remain either an odd number of quarks or an odd number of antiquarks between the two shown in Fig.~\ref{fig:sym}. 
These cannot possibly be joined up in a planar way, and it follows that there can be no such scalar contributions.

The flavour recursion proceeds in a similar fashion to the one described in Sec.~\ref{flav}. Since the direction of
the quark lines alternate between levels in the Dyck tree, now two flavours must
be used to stop the flavour mixing between odd and even levels of the Dyck tree. That is, the distinct flavour primitive
in the alternating basis is expressed as a two-flavour primitive of the same labelling, with one flavour, $f_1$, for quarks
corresponding to odd levels of the Dyck tree, and the other flavour $f_2$ assigned to even levels of the Dyck tree -- see rhs
of Fig.~\ref{fig:twosym} -- minus distinct flavour subtraction primitives to subtract the wrong flavour running. It is clear
that flavour paring  cannot occur between odd and even levels. Considering now just the odd
levels, the same arguments as explained in Sec.~\ref{flav} (see Fig.~\ref{fig:flavmix}) apply to these quark lines of flavour $f_1$,
that is, the non zero permutations only occur at the same level of the Dyck tree. 
The same applies to quark lines at even levels. Each of the subtraction
primitives is of higher maturity, and needs to be re-expressed in terms of primitives of alternating basis form. In this
way, the recursion terminates at primitives of the form
\be
\ma(1_{f_1} \, q_{f_2} \, \bar{q}_{f_1} \, q_{f_2} \, \bar{q}_{f_1} \, \ldots, q_{f_1} \, \bar{q}_{f_2} \,  q_{f_1} \, \bar{q}_{f_2} \, 2_{f_1})\,.
\ee
These are the primitives which have a Dyck tree of highest maturity (shown on the rhs of Fig.~\ref{fig:maxheight}).

\bibliography{getmoreflav.bib}

\end{document}